\documentclass[aps,prb,amsmath,groupedaddress,letterpaper,showpacs,twocolumn]{revtex4}
\usepackage{graphicx,array,tabularx}
\usepackage{bm,amsmath,amssymb,verbatim,slashbox}

\begin{document}

\title{Model for twin electromagnons and magnetically induced oscillatory
  polarization in multiferroic {\it R}MnO$_3$}

\author{Markku P. V. Stenberg}
\email[]{markku.stenberg@iki.fi}
\author{Rog\'{e}rio de Sousa}
\email[]{rdesousa@uvic.ca}
\affiliation{Department of Physics and Astronomy, University of Victoria,
Victoria, British Columbia, Canada V8W 3P6}

\begin{abstract}
  We propose a model for the pair of electromagnon excitations
  observed in the class of multiferroic materials {\it R}MnO$_3$ 
  ({\it R} is a
  rare-earth ion). The model is based on a harmonic cycloid ground
  state interacting with a zone-edge magnon and its twin excitation
  separated in momentum space by two times the cycloid wave vector.
  The pair of electromagnons is activated by cross coupling between
  magnetostriction and spin-orbit interactions. Remarkably, the spectral 
  weight of the twin electromagnon is
  directly related to the presence of a magnetically induced
  oscillatory polarization in the ground state. This leads to the
  surprising prediction that TbMnO$_3$ has an oscillatory polarization
  with amplitude $50$ times larger than its uniform polarization.
\end{abstract}
\date{\today}

\maketitle

\section{Introduction}

The coexistence of magnetic and ferroelectric phases in multiferroic
materials gives rise to important effects related to
cross correlation between order parameters and external
fields.\cite{kimura03,smolenskii82} 
A notable consequence is that the
elementary excitations are not purely magnetic or ferroelectric in
character. The spin waves are admixed with vibrational modes of the
electric polarization, giving rise to electric dipole active
magnons, the so-called electromagnons. 
Electromagnons were postulated
to exist in 1969 by Bary'akhtar and
Chupis \cite{baryakhtar} but their existence was
confirmed only recently by sensitive optical experiments in the far
infrared\cite{pimenov06,sushkov07,cazayous08,kida08b,aguilar09,lee09}
combined with neutron scattering.\cite{senff07}

Apart from its fundamental importance, the electromagnon spectra gives
invaluable information on how magnetism couples to ferroelectricity.
The class of perovskite manganites {\it R}MnO$_3$
(RMO) is playing a major role in this respect. Here {\it R} is a rare-earth
ion such as Gd, Tb, Dy, or mixtures between them (e.g.,
Gd$_{x}$Tb$_{1-x}$, etc.). At low temperatures ($T\sim 20$~K)
the Mn spins are ordered in a spiral state with period
incommensurate with the lattice.\cite{kenzelmann05b} The spin spiral is
of the cycloid type. This breaks space inversion and gives rise to a
ferroelectric moment along one of the directions in the cycloid
plane.\cite{mostovoy06} At the microscopic level, this ferroelectric
moment originates from Dzyaloshinskii-Moriya (DM)
coupling.\cite{katsura05,mostovoy06,sergienko06,malashevich08}

Nevertheless, contrary to predictions based on DM
coupling,\cite{katsura07} electromagnons in RMO are observed only when
the electric field of light is along the crystallographic direction
$\hat{a}$.\cite{kida08b,aguilar09,pimenov09,lee09} Moreover, for all
ions {\it R}, \emph{two electromagnons are always
  observed}.\cite{note_thirdEM} Recently, the origin of one of these
electromagnons was explained by Aguilar {\it et al.},\cite{aguilar09,lee09} 
who pointed out that the high frequency electromagnon was a zone-edge magnon 
activated by pure magnetostriction. However, the origin of
the low-frequency electromagnon remains unknown.

A recent theory\cite{desousa08} shows that the anharmonicity of a
distorted cycloid ground state gives rise to multiple electromagnons
that were observed using Raman scattering in
BiFeO$_3$.\cite{cazayous08} However, the anharmonicity mechanism 
cannot explain the pair of electromagnons in RMO because the degree of
anharmonicity detected by neutron diffraction is too low in these
materials.\cite{kenzelmann05b}

Here we propose a model that is able to explain the double
electromagnon feature observed in RMO.  We show that a simple
\emph{harmonic} cycloid ground state with wave vector $Q$ has a special
pair of twin electromagnon excitations, located at the zone edge and
$2Q$ away from it.  Remarkably, the activation of the latter
electromagnon is directly related to the presence of an oscillatory
polarization with wave vector $2Q$ in the ground state.

\section{Model}

\begin{figure}
\includegraphics[width = 3 in]{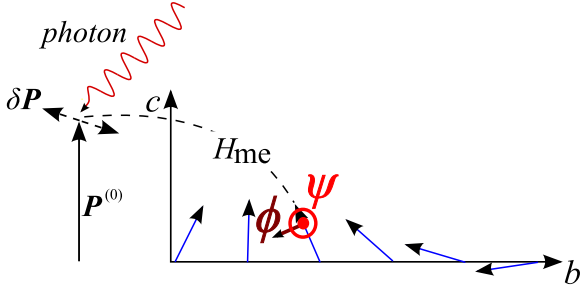}
\caption{(Color online) Coupled spin waves and optical phonons in the
  cycloid ground state. In-plane spin fluctuations are denoted by
  $\phi$, while out-of-plane spin fluctuations are denoted by $\psi$.}
\label{fig:gs_and_fluctuations}
\end{figure}

The usual phenomenological Landau theory is based on a free-energy
expansion into powers of the spatial derivatives of the order
parameters.\cite{lifshitz80} Therefore, the 
Landau approach\cite{baryakhtar,mostovoy06,desousa08,glinchuk08} 
cannot describe large wave vector excitations such as zone-edge magnons.
Here we adopt a model that is more microscopic than the 
Landau theory. Our model Hamiltonian consists of
three distinct contributions, $H=H_{S}+H_{\rm{ph}}+H_{\rm{me}}$. The
first contribution describes spin frustration in
RMO,\cite{aguilar09,lee09,senff07,kenzelmann05b,kajimoto05}
\begin{equation}
H_{S} = \sum_{n,m}J_{n,m}\hat{{\bf S}}_{n}\cdot\hat{{\bf S}}_{m}
+D\sum_{n}(\hat{{\bf S}}_{n}\cdot \hat{a})^{2}.
\label{eq:H_S}
\end{equation}
Here $J_{n,m}$ is the exchange coupling between Mn spins
at positions ${\bm R}_{n}$ and ${\bm R}_{m}$, and $D$ is a single-ion
anisotropy along the crystallographic direction $\hat{a}$ (we assume
$D>0$, i.e., easy-plane anisotropy). RMO has an orthorhombic lattice,
with an orthogonal system of crystal axis $\hat{a},\hat{b},\hat{c}$
and unequal bond lengths $a\neq b\neq c$ .  There are four Mn spins
per unit cell; spins 1 and 2 lie in the {\it ab} planes, with spins 3 and 4
a distance $c/2$ above them. The coupling between nearest neighbor
spins in the {\it ab} planes is ferromagnetic, and will be denoted by $J_0$;
the coupling between {\it ab} layers along the $\hat{c}$ direction is
antiferromagnetic and is denoted by $J_c$. Spin frustration arises due to
coupling between next-nearest neighbors along the $\hat{b}$ direction;
this is antiferromagnetic and denoted by $J_{2b}$. When the stability
condition $J_{2b}>-J_0/2$ is satisfied, the ground state is a {\it bc}
cycloid\cite{katsura05,sergienko06,malashevich08} as observed in
experiments,\cite{kimura03,kenzelmann05b}
\begin{equation}
\bm{S}_{0}({\bf R}) =
\pm
S[\cos({\bf Q}\cdot {\bf R})\hat{b}+
\sin({\bf Q}\cdot {\bf R})\hat{c}].
\label{groundstate}
\end{equation}
Here $\bm{Q}=Q\hat{b}$ is the cycloid wave vector, with
$\cos{\left(Qb/2\right)}=-J_0/(2J_{2b})$.  The upper (+) sign applies
to {\it ab} layer spins $1$ and $2$, whereas the lower (--) sign applies to
spins $3$ and $4$ in the neighboring {\it ab} layers immediately above and 
below them. Figure~\ref{fig:gs_and_fluctuations} depicts the cycloid ground
state.
The second ingredient of our model is the lattice fluctuation Hamiltonian,
\begin{equation}
H_{\rm{ph}}=\frac{1}{2}m^{*}\sum_n \left(\dot{\bm{x}}_{n}^{2}
+\omega_{0}^{2} \bm{x}_{n}^{2}\right)
-e^{*}\sum_n \bm{x}_n \cdot \bm{E},
\end{equation}
where $\bm{x}_{n}$ is a relative displacement between cations and
anions within each unit cell. Here $e^{*}$ is the Born effective
charge, $m^{*}$ is an effective mass, and $\omega_{0}$ is the (bare)
phonon frequency.  The modes $\bm{x}_{n}$ are directly related to the
polarization order parameter in Fourier space,
$\bm{P}_{\bm{q}}(\omega)=\frac{e^{*}}{N v_0}\int\mathrm{d}t\sum_{n}
  \textrm{e}^{i
  (\bm{q}\cdot \bm{R}_{n}-\omega t)}\bm{x}_{n}$, where $v_0$ is the
volume of the unit cell.  The electric field of light,
$\bm{E}\textrm{e}^{-i\omega t}$, couples linearly to the $q=0$
polarization, $\bm{P}_{0}(\omega)=\chi_{0}(\omega)\bm{E}$, with
electric susceptibility $\chi_0
(\omega)=(e^{*2})/[(m^{*}v_0)(\omega_{0}^{2}-\omega^{2})]$.

The third contribution $H_{\rm{me}}$ models the coupling between the
spin degrees of freedom and the lattice. Following the experimental
observation that electromagnons are only excited when the electric
field is along $\hat{a}$,\cite{kida08b,aguilar09,pimenov09,lee09} we
consider the interaction,
\begin{eqnarray}
H_{\rm{me}} &=&
e^{*}\sum_{n}x_{n}^{a}[
g_{c}(\hat{S}_{1,n}^{c}-\hat{S}_{1,n+b}^{c})(\hat{S}_{2,n}^{c}
+\hat{S}_{2,n+a}^{c})\nonumber\\& &+g_{b}(\hat{S}_{1,n}^{b}
-\hat{S}_{1,n+b}^{b})(\hat{S}_{2,n}^{b}
+\hat{S}_{2,n+a}^{b})\nonumber\\& &+(1\rightarrow 3, 2\rightarrow 4)],
\label{eq:linear_pbnm}
\end{eqnarray}
which is invariant under the {\it Pbnm} space-group operations of RMO.  Here
$g_c\neq g_b$ are coupling constants with dimension of electric
fields; we neglect the components proportional to $g_a$ because these
play no role in the discussion below.  Remarkably,
Eq.~(\ref{eq:linear_pbnm}) is \emph{not} rotationally invariant in
spin operators. Therefore it differs in an important way from the pure
magnetostrictive coupling considered previously.\cite{aguilar09} \emph{The
absence of rotational invariance in magnetoelectric coupling is a
direct consequence of cross coupling between magnetostriction and
spin-orbit interactions, and should be common property of all
multiferroics with orthorhombic lattice} (this follows from symmetry
since $a\neq b\neq c$ precludes rotations that take one axis into the
other). Later we will explain why the experimental observations favor
this coupling over other symmetry-allowed explanations.

It turns out that Eq.~(\ref{eq:linear_pbnm}) gives rise to a remarkable static 
effect. Minimizing $H_{\rm{ph}}$ with respect to $\bm{x}_n$, and plugging in 
the cycloidal spin order we get
\begin{equation}
\frac{e^*\bm{x}_{n}}{v_0}=P_{\rm{IOP}}\sin{[(2n+1)Qb]}\hat{a}.
\end{equation}
Hence the local polarization per unit cell is oscillatory with
wave vector $2Q$; the amplitude for the \emph{incommensurate oscillatory
polarization} (IOP) is given by
\begin{equation}
P_{\rm{IOP}}=4\left(g_b-g_c\right)
\chi_0 S^2
\sin{\left(\frac{Qb}{2}\right)}.
\label{PIOP}
\end{equation}
Remarkably, the presence of the IOP is directly related to the lack of
rotational invariance in Eq.~(\ref{eq:linear_pbnm}), i.e., the fact
that $g_b\neq g_c$.

It is well known that TbMnO$_3$ has a magnetically-induced lattice
modulation with wave vector $2Q$ in the
cycloidal phase.\cite{kimura03} However, we are not aware of any
claims that such a modulation leads to oscillatory electric
polarization. The main result of this paper is to show that this
oscillatory polarization is directly related to the spectral weight of
the twin electromagnon; hence it can be detected by optical
experiments.

\section{Twin electromagnon excitations}

We now consider the elementary excitations of the coupled spin-phonon
system. The spin excitations $\delta
\hat{\bm{S}}=\hat{\bm{S}}-\bm{S}_{0}$ are parametrized as follows,
\begin{equation}
\delta \hat{\bm{S}}_{i,n}= \hat{\psi}_{i,n} \hat{a}
\pm \hat{\phi}_{i,n}[
\cos{(\bm{Q}\cdot \bm{R}_{i,n})}\hat{c}
-\sin{(\bm{Q}\cdot \bm{R}_{i,n})}\hat{b}],
\end{equation}
where the sign convention is the same as in Eq.~(\ref{groundstate}). Here the
operators $\hat{\psi}$ describe spin fluctuations out of the cycloidal
plane, and $\hat{\phi}$ denotes in-plane (tangential) fluctuations
(Fig.~\ref{fig:gs_and_fluctuations}).  We carry out a mean-field
expansion of the Hamiltonian $H$ about its equilibrium value
by keeping only terms quadratic in
the fluctuation operators, e.g., 
$\delta P^{2}$, $\hat{\phi}^2$, $\delta P\hat{\psi}$, etc.
Using the canonical commutation relations
$[\hat{\phi}_{j,n},\hat{\psi}_{k,m}]=i\delta_{jk}\delta_{nm}S$,
we derive the coupled equations of motion for spin and polarization,
\begin{subequations}
\begin{eqnarray}
&\left(\omega^{2}-\Omega_{\rm{C},q}^{2}\right)
\left(\phi_{1q}+\phi_{2q}+\phi_{3q}+\phi_{4q}\right)=\Gamma_{q-k_0},
\label{cyclon}
\\
&\left(\omega^{2}-\Omega_{\rm{C},q+k_0}^{2}\right)
\left(\phi_{1q}-\phi_{2q}+\phi_{3q}-\phi_{4q}\right)=\Gamma_q,
\label{zecyclon}
\\
&\left(\omega^{2}-\Omega_{\rm{EC},q}^{2}\right)
\left(\phi_{1q}+\phi_{2q}-\phi_{3q}-\phi_{4q}\right)=0,
\label{extracyclon}
\\
&\left(\omega^{2}-\Omega_{\rm{EC},q+k_0}^{2}\right)
\left(\phi_{1q}-\phi_{2q}-\phi_{3q}+\phi_{4q}\right)=0.\label{zeextracyclon}
\end{eqnarray}
\end{subequations}
Here $\phi_{iq}$ are Fourier transforms of the fields $\phi_{in}$;
identical equations hold for the fields
$\psi_{in}$.\cite{note_relation_psi_phi} There are four spin-wave
modes, each transforming according to four different 1{\it d}
representations of the {\it Pbnm} space-group.  The second mode
[Eq.~(\ref{zecyclon})] is related to the first mode by
$\bm{q}\rightarrow \bm{q}+\bm{k}_{0}$, with
$\bm{k}_{0}=(2\pi/b)\hat{b}$ the Brillouin zone edge (the lattice has
periodicity $b/2$ along $\hat{b}$). This corresponds to the fact that
``anti-phase'' fluctuations of the neighboring spins with the wave
vector $q$ are equivalent to ``in-phase'' fluctuations at $q+2\pi/b$.
Their dispersion satisfies $\Omega_{\rm{C},q\rightarrow 0}\rightarrow 0$,
reflecting the phase sliding symmetry of the cycloid. Adding a
constant phase $\phi$ to Eq.~(\ref{groundstate}) yields a different
ground state with the same energy.  At $q=0$ the first mode is a pure
phase fluctuation ($\phi_{in}$ is the same for all $i$ and
$\psi_{in}=0$). For this reason this mode will be referred as a
\emph{cyclon}, with dispersion $\Omega_{\rm{C},q}$. The second mode
is naturally called a \emph{zone-edge cyclon}.
\begin{figure}
\includegraphics[width = 3 in]{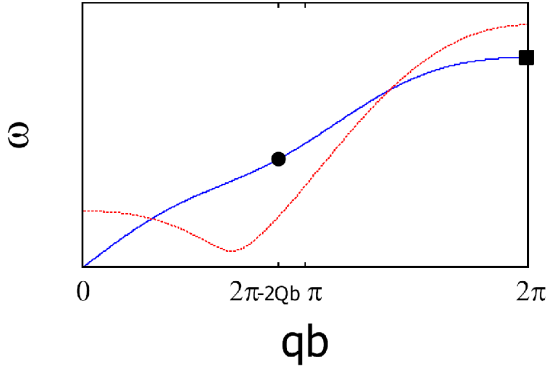}
\caption{(Color online) Typical dispersion curves along the cycloid
  direction $\hat{b}$ for the cyclon (blue solid) and extra-cyclon
  (red dashed) modes in a {\it bc} cycloid. Our theory gives rise to twin
  electromagnons depicted by a circle and a square. When the
  magnetoelectric interaction is rotationally invariant [$g_b=g_c$ in
  Eq.~(\ref{eq:linear_pbnm})], only the one denoted by a square is
  excited; in the absence of rotational invariance, both
  electromagnons are excited. Here $Q$ is the magnitude of the cyclon
  wave vector and $b$ the lattice constant along $\hat{b}$ axis.}
\label{fig:edge_twin}
\end{figure}
Similarly, the fourth mode can be obtained from the third by the same
translation in {\it q} space, $\bm{q}\rightarrow \bm{q}+\bm{k}_{0}$.
However, the third and fourth modes are gapped (gap proportional to $J_c$). 
These modes will be denoted \emph{extra cyclons}. The (bare)
magnon dispersions are shown in Fig.~\ref{fig:edge_twin}.  For
$\bm{q}$ along $\hat{b}$ they are given by
\begin{subequations}
\begin{eqnarray}
  \Omega_{\rm{C},q}^2 &=& \frac{4S^2}{\hbar^2}\left[\left(J_0+2J_{2b}\cos{\frac{qb}{2}}\right)^{2}+2J_{2b}(D+2J_c)\right]\nonumber\\
&&\times\left(4\cos^{2}{\frac{qb}{4}}-\frac{J_{0}^{2}}{J_{2b}^{2}}\cos{\frac{qb}{2}}\right)
  \sin^{2}{\frac{qb}{4}},\label{cyclon_spectra}\\
\Omega_{\rm{EC},q}^{2} &=& \frac{4 S^2}{\hbar^2} \left[\left(J_0+2J_{2b}\cos{\frac{qb}{2}}\right)^{2}+2J_{2b}D\right]\nonumber\\
&\times&\left[\left(4\cos^{2}{\frac{qb}{4}}-\frac{J_{0}^{2}}{J_{2b}^{2}}\cos{\frac{qb}{2}}\right)
 \sin^{2}{\frac{qb}{4}}+\frac{J_c}{J_{2b}}\right].\label{extracyclon_spectra}
\end{eqnarray}
\end{subequations}
Typical values for the exchange couplings are $J_0\sim -0.4$ meV,
$J_{2b}\sim 0.3-0.4$ meV, and $J_{c}\sim 0.5-1$ meV. \cite{lee09}
Interestingly, only the zone-edge cyclon is coupled to the
polarization. This occurs through the dynamical magnetoelectric field,
\begin{eqnarray}
\Gamma_q &=& \frac{8 v_0}{\hbar^2}S^{2}\sin{\left(\frac{Qb}{2}\right)}\left\{
2J_0 \left[\cos{\left(\frac{qb}{2}\right)}+\cos{\left(\frac{Qb}{2}\right)}\right]\right.\nonumber\\
&&\left.-J_{2b}\left[\cos{\left(qb\right)}-\cos{\left(Qb\right)}\right]-(D+2 J_c)\right\}\nonumber\\
&&\times\left\{
2(g_b+g_c)P_{q}^{a}(\omega)\right.\nonumber\\
&&\left.-(g_b-g_c)\left[P_{q-2Q}^{a}(\omega)+P_{q+2Q}^{a}(\omega)\right]
\right\}.
\end{eqnarray}
Similarly, the equation of motion for the polarization only couples
$P_{0}^{a}$ to the zone-edge cyclon at $q=0$ and $q=\pm 2Q$.
\emph{Hence we have a set of twin electromagnons at wave vectors
  $q=k_0$ and $q=k_0-2Q$}.  For $g_b=g_c$, only one electromagnon (the
zone-edge cyclon) is activated.\cite{aguilar09} This excitation
corresponds to a scattering process with in and out-going momenta
equal to the cycloid wave vector $Q$
[Fig.~\ref{fig:mom_transfer_0and2Q}(a)]. However, when $g_b\neq g_c$,
the twin electromagnon (a cyclon at $q=k_0-2Q$) appears, corresponding
to two outgoing momenta adding up to $2Q$
[Fig.~\ref{fig:mom_transfer_0and2Q}(b)].
\begin{figure}
\includegraphics[width = 3 in]{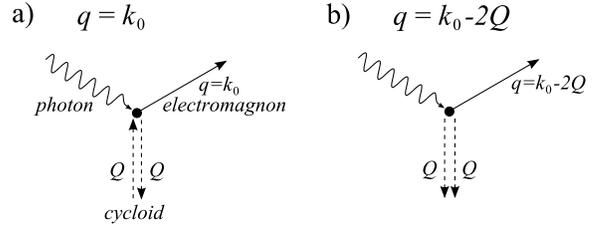}
\caption{Diagrams for twin electromagnon excitation by
  light, involving momentum exchange with the cycloidal ground state.
  (a) Zone-edge electromagnon at $q=k_0$ and (b) its twin at
  $q=k_0-2Q$.  The fact that momentum is not conserved by $k_0=2\pi/b$
  reflects the underlying lattice symmetry, which produces spin-wave
  modes connected by $q\rightarrow q+k_0$.}
\label{fig:mom_transfer_0and2Q}
\end{figure}
\section{Dielectric function and comparison to experiments}

In order to relate to optical experiments, we calculate the dielectric
function $\varepsilon(\omega) = \varepsilon_\infty +4\pi
P_{0}^{a}(\omega)/E^{a}$ and study its resonances. When the resonance
frequencies are not too close to each other $\varepsilon$ is a sum of 
Lorentzians,
\begin{eqnarray}
\varepsilon(\omega)&=&\frac{{\cal S}_{k_0}}{\Omega_{\rm{C},k_0}^2-\Delta_{k_0}^2-\omega^2}
+\frac{{\cal S}_{k_0-2Q}}{\Omega_{\rm{C},k_0-2Q}^2-\Delta_{k_0-2Q}^2-\omega^2}\nonumber\\
&&+\frac{{\cal S}_{\rm{ph}}}{\omega^2_0+\Delta_{\rm{ph}}^{2}-\omega^2}+\varepsilon_{\infty}.
\end{eqnarray}
The electromagnon frequencies are seen to be downshifted from the bare
magnon frequency $\Omega_{\rm{C},q}$ to
$\sqrt{\Omega_{\rm{C},q}^{2}-\Delta_{q}^{2}}$, while the phonon
frequency $\omega_0$ gets upshifted to $\sqrt{\omega_{0}^{2}+
  \Delta_{\rm{ph}}^{2}}$.  The spectral weights ${\cal S}_{q}$ share a simple
relationship with the frequency shifts: For electromagnons, ${\cal
  S}_{q}= [\varepsilon(0)-\varepsilon_\infty]\Delta_{q}^{2}$, while for the 
phonon 
${\cal  S}_{\rm{ph}}=[\varepsilon(0)-\varepsilon_\infty](\omega_{0}^{2}-\Delta_{\rm{ph}}^{2})$.
The frequency shifts are related by
$\Delta_{\rm{ph}}^{2}=\Delta_{k_{0}}^{2}+\Delta_{k_0-2Q}^{2}$,
ensuring the satisfaction of the oscillator
strength sum rule, ${\cal S}_{k_0}+{\cal S}_{k_0-2Q}+{\cal
  S}_{\rm{ph}}=\omega_{0}^{2}\chi_{0}$.
The electromagnon spectral weights are given by
\begin{subequations}
\begin{eqnarray}
\label{Sk0}
&{\cal S}_{k_0}=\frac{4\pi S \chi_{0}^{2} v_0\omega_0^4
(g_b+g_c)^{2}}{(\omega_0^2-\Omega_{c,k_{0}}^2)^2}
\tan^{2}{\frac{Qb}{2}}
\left(\frac{\Omega_{\rm{C},k_0}^{2}}{J_{2b}}\right),
\\
&{\cal S}_{k_0-2Q}=
\frac{
2\pi S \chi_{0}^{2} v_0\omega_0^4(g_b-g_c)^{2}
\tan^{2}{\frac{Qb}{2}}}
{(\omega_0^2-\Omega_{c,k_{0}-2Q}^2)^2
\left[\cos^{4}{\frac{Qb}{2}}+\sin^{4}{\frac{Qb}{2}}\right]}
\left(\frac{\Omega_{\rm{C},k_0-2Q}^{2}}{J_{2b}}\right).\nonumber\\
\label{Sk0m2Q}
\end{eqnarray}
\end{subequations}

\begin{widetext}
\begin{center}
\begin{table}
\begin{center}
\scalebox{0.94}{
  \begin{tabular}{ c  c  c  c  c   c}
    \hline
    \hline
\hspace{3mm}
&\hspace{3mm}Gd$_{0.5}$Tb$_{0.5}$\hspace{3mm}
&\hspace{3mm}Gd$_{0.1}$Tb$_{0.9}$\hspace{3mm}
&\hspace{3mm}Tb\hspace{3mm}
&\hspace{3mm}Tb$_{0.41}$Dy$_{0.59}$\hspace{3mm}
&\hspace{3mm}Dy\hspace{3mm}
\\
\hline
$\left(\frac{\Omega_{A}}{\Omega_{B}}\right)_{\mathrm{Exp}}$
& 0.30
& 0.39
& 0.40
& 0.36
& 0.37\\
$\left(\frac{\Omega_{c,k_0-2Q}}{\Omega_{c,k_0}}\right)_{\mathrm{Th}}$
& 0.55
& 0.53
& 0.52
& 0.52
& 0.54\\
$\left(\frac{\mathcal{S}_{A}}{\mathcal{S}_{B}}\right)_{\mathrm{Exp}}$
& 0.08
& 0.11
& 0.13
& 0.41
& 0.51\\
$\left[\frac{(g_b+g_c)^2}{(g_b-g_c)^2}\frac{\mathcal{S}_{c,k_0-2Q}}{\mathcal{S}_{c,k_0}}\right]_{\mathrm{Th}}$
& 0.14
& 0.16
& 0.16
& 0.16
& 0.18
\\ \hline
    \hline
  \end{tabular}}
\caption{This table compares our theoretical calculations to
  experiments in {\it R}MnO$_3$. A and B are, respectively,
  the lower- and higher-energy electromagnon modes detected in
  [\onlinecite{lee09}].}
\label{tb:freq_sw}
\end{center}
\end{table}
\end{center}
\end{widetext}
An important experimental result is that the  
spectral weights are nearly the same for the {\it ab} and {\it bc} cycloids.
\cite{kida08b,aguilar09,pimenov09}
It is easy to see that this result follows from our model when
$|g_b|\gg |g_a|, |g_c|$.
 Note that in this limit the magnetoelectric
coupling Eq.~(\ref{eq:linear_pbnm}) is invariant under the flip of the
cycloid plane, in spite of the fact that it lacks rotational symmetry.
Table I compares our theoretical calculations to the measured values
$\Omega_A/\Omega_B$ and ${\cal S}_A/{\cal S}_B$, where A and B label
the lower- and higher-energy electromagnons observed in
RMO.\cite{lee09} We used the same model parameters as the ones in 
Fig.~4(b) of Ref.~[\onlinecite{lee09}] (filled symbols).  The variations
in the ratio $\mathcal{S}_{A}/\mathcal{S}_{B}$ indicate differences in
the couplings $g_c,g_b$ for different ions {\it R}.  For Tb our theory
matches the observed values when $g_c/g_b=0.05$; for Dy we get
$g_c/g_b=-0.25$.

\section{Discussion and conclusions}
We now consider other possibilities for the activation of the 
low-frequency electromagnon. Interactions containing crossed terms such as
$x_{n}^{a}\hat{S}_{i}^{b}\hat{S}_{j}^{c}$ are also allowed by the {\it Pbnm}
symmetry. For a {\it bc} cycloid, this interaction gives rise to an
electromagnon at $q=k_0-2Q$. However, when the cycloid is flipped to
the {\it ab} plane, this interaction leads instead to an electromagnon at
$q=k_0-Q$, with no electromagnon at $q=k_0-2Q$. This result implies a
large shift in electromagnon frequency (more than 50\% as seen in
Fig.~\ref{fig:edge_twin}), that is in contradiction to
experiments.\cite{kida08b,aguilar09,lee09,pimenov09} Therefore the
interaction $x_{n}^{a}\hat{S}_{i}^{b}\hat{S}_{j}^{c}$ can not explain
the origin of the low-frequency electromagnon. Similarly, interactions
of the form $x_{n}^{a}\hat{S}_{i}^{a}\hat{S}_{j}^{c}$ only give rise
to electromagnons at $k_0-Q$; this leads to ratios of electromagnon
frequencies $\Omega_A/\Omega_B$ 
that are twice as large as obtained experimentally (Table I).\cite{lee09} 
Hence $x_{n}^{a}\hat{S}_{i}^{a}\hat{S}_{j}^{c}$ is also ruled out.

Aguilar {\it et al.}\cite{aguilar09} suggested that an elliptical
spiral structure such as the one in
BiFeO$_3$ (Refs.~\onlinecite{cazayous08} and \onlinecite{desousa08}) 
may explain the low-frequency
electromagnon. This possibility is also ruled out because the
degree of ellipticity observed in neutron-scattering
experiments\cite{kenzelmann05b} is too low to explain the large
spectral weight of the low frequency electromagnon. Furthermore,
Ref.~\onlinecite{aguilar09} suggested that purely magnetic
interactions such as $\hat{S}_{i}^{a}\hat{S}_{j}^{c}$ would mix
extra-cyclon magnons at $q=Q$ to the zone-edge electromagnon,
providing an alternative explanation for the low frequency resonance.
However, the extra-cyclon magnon energy at $q=Q$ is
approximately constant for different {\it R}'s, in contradiction to the
trend observed in experiments [$\Omega_A$ tends to decrease with
increasing ionic radius; $\Omega_A$ for Gd$_{0.5}$Tb$_{0.5}$ is 
40\% higher than for Dy (Ref.~\onlinecite{lee09})]. Interestingly, our theory gives a
natural explanation for this trend since the value of the cyclon
energy at $q=k_0-2Q$ also decreases appreciably as the ionic radius
increases.

We now discuss the implications of our model for the understanding of
multiferroic order.  One remarkable consequence of the presence of the
$q=k_0-2Q$ electromagnon is that its spectral weight can be directly
related to the presence of the 
IOP in the ground state. Using Eqs.~(\ref{PIOP})~and~(\ref{Sk0m2Q})
we derive an important relation between the amplitude of the IOP and
the twin electromagnon spectral weight,
\begin{equation}
P_{\rm{IOP}}^{2}=\frac{8S^3 J_{2b}}{\pi v_0}
\cos^{2}{\frac{Qb}{2}}\left[\cos^{4}{\frac{Qb}{2}}
+\sin^{4}{\frac{Qb}{2}}\right]
\frac{{\cal S}_{k_0-2Q}}{\Omega_{k_0-2Q}^{2}}.
\label{PIOPSrelation}
\end{equation}
Hence optical experiments are a direct probe of the IOP amplitude.
For TbMnO$_3$, the measured twin electromagnon frequency is
$\Omega_{\rm{C},k_0-2Q}=25$~cm$^{-1}$, with spectral weight ${\cal
  S}_{k_0-2Q}=1.7\times 10^{3}$~cm$^{-2}$.\cite{lee09}
Using
Eq.~(\ref{PIOPSrelation}) we obtain $P_{\rm{IOP}}=4$~$\mu$C/cm$^{2}$,
a value that is $50$ times larger than the uniform polarization $P_0
=8\times 10^{-2}$~$\mu$C/cm$^{2}$ present in TbMnO$_3$.

In conclusion, we introduced the concept of the twin electromagnon in
order to explain optical experiments in the {\it R}MnO$_3$ family of
multiferroics.  Our symmetry analysis shows that an incommensurate
oscillatory polarization coexists with the well-known cycloid phase in
these materials. Remarkably, there exists a direct relation between
the twin electromagnon spectral weight and the amplitude of this
oscillatory polarization. Hence we showed that TbMnO$_3$ has an
oscillatory polarization with amplitude 50 times larger than its
uniform polarization.  This surprising conclusion underlines the
importance of electromagnons in the characterization of multiferroic
order.
\section{Acknowledgments}
We thank N. Kida and A. Pimenov for useful discussions. This research
was supported by NSERC discovery and the UVic Faculty of Sciences.


\begin{thebibliography}{23}
\bibitem[{A third, lower frequency electromagnon with quite small spectral
  weight was reported in \onlinecite{pimenov09}.()}]{note_thirdEM}
A third, lower frequency electromagnon with quite small spectral weight was
  reported in \onlinecite{pimenov09}.
\bibitem[{There exists a linear relation connecting the fields $\psi_{in}$ to
  the fields $\phi_{in}$; therefore the $\psi_{in}$ are completely specified
  once the $\phi_{in}$ are set.()}]{note_relation_psi_phi}
There exists a linear relation connecting the fields $\psi_{in}$ to the fields
  $\phi_{in}$; therefore the $\psi_{in}$ are completely specified once the
  $\phi_{in}$ are set.
\end{thebibliography}
\end{document}